\documentclass[epj]{svjour}
\usepackage{graphics}

\newcommand{\beq}{\begin{equation}}
\newcommand{\eeq}{\end{equation}}
\newcommand{\beqas}{\begin{eqnarray*}}
\newcommand{\eeqas}{\end{eqnarray*}}
\newcommand{\beqar}{\begin{eqnarray}}
\newcommand{\eeqar}{\end{eqnarray}}
\newcommand{\req}[1]{(\ref{#1})}

\begin{document}
\title{Self Organization of Interacting Polya Urns}
\author{Matteo Marsili\inst{1} \and Angelo Valleriani\inst{2}}
\institute{International School for Advanced Studies (SISSA)
and INFM unit, V. Beirut 2-4, I-34014, Trieste \and
Max-Planck-Institut f\"ur Physik Komplexer 
Systeme,
N\"othnitzer Str. 38, D-01187 Dresden, Germany}
\date{\today}
\abstract{
We introduce a simple model which shows non-trivial self organized
critical properties. The model describes a system of interacting units, 
modelled by Polya urns, subject to perturbations and which occasionally
break down. Three equivalent formulations - stochastic, quenched and 
deterministic - are shown to reproduce the same dynamics. Among
the novel features of the model are a non-homogeneous stationary
state, the presence of a non-stationary critical phase and non-trivial
exponents even in mean field. We discuss simple interpretations 
in term of biological evolution and earthquake dynamics and we
report on extensive numerical simulations in dimensions
$d=1,2$ as well as in the random neighbors limit.
}
\maketitle

Our understanding of Self Organized Criticality (SOC) \cite{btw},
as a general framework for the emergence of scale-free behavior
in Nature, has greatly benefitted from the introduction of
simple models. Even though models such as the
sandpile \cite{btw} and the Bak-Sneppen \cite{bs}
are too simple to capture the complexity of natural
phenomena such as earthquakes\cite{earthq} and biological 
evolution\cite{Gould89,fossil}, they have, nonetheless, identified
some basic mechanisms leading to SOC. These systems have been 
a starting point both for the development of more complex and
realistic models of natural phenomena\cite{modbtw,modbs}, 
and for analytical approaches\cite{dhar,RSRG,PMB,mf}, which have 
led us to a much deeper understanding of SOC. Indeed,
we can now identify some basic ``routes to
Self Organized Criticality'' such as those based on
sandpile \cite{btw}, extremal dynamics\cite{PMB,RTS},
memory\cite{memo} and network \cite{netw} models.

In this Rapid Communication we propose a qualitatively different
``route to SOC'' based on a very simple model of
interacting Polya urns. 
Its qualitative differences with respect to other
SOC models are that it is characterized by a 
non-homogeneous stationary state and by non 
trivial exponents even in the mean field case.
Furthermore, we shall show numerical evidence
for the occurrence of a {\em non-stationary 
self organized critical state}. Moreover, the model can be formulated in three
different but equivalent ways. 
This fact, on one hand allows us to use a wide variety of 
tools to investigate its critical properties, and on
the other it bridges different descriptions of 
the same process. All these features can well
be relevant in the description of natural phenomena.
The model indeed provides a general framework
for the emergence of SOC which, as we shall
discuss, can be applied both to coevolution
and to large scale earthquakes dynamics.
Note indeed that the patterns of earthquakes
activity are highly non-homogeneous and that such
a system is, in principle, non-stationary. 
The same applies to our ecosystem, which is in a non-stationary state where
ever fitter species replace less fit ones.

We consider a system of interacting 
Polya urns arranged on a $d$-dimensional lattice.
A Polya urn is a simple model to study e.g.\  the
occurrence of accidents\cite{feller}.
Each urn contains initially $b$ black balls and
$1$ white one. As in sandpile models, at each time step
we randomly select a site and attempt to add a ``grain of
sand'', i.e.\ a white ball, to the corresponding urn. 
A ball is drawn from the selected urn: 
If the ball is white the attempt is successful and a 
new white ball is added to the urn. If it is black 
a ``fatal accident'' occurs: 
The urn becomes unstable and it ``topples''. 
The toppling mechanism is as follows:  
{\em 1)} the urn is reset to $1$ white ball and $b$ 
black ones and {\em 2)} for each  
white ball of the original urn a similar attempt
to add a white ball is made on 
a randomly chosen nearest neighbor urn.
In this way, white balls released by an unstable urn
can provoke some ``fatal accident'' in nearby urns
(addition of white ball to already
unstable urns leaves them unstable but it increases the number
of white balls in it). The process stops when all
balls are redistributed provoking no further toppling.
A new attempt to add a ball to a randomly 
chosen urn is made, at the next time-step, and the process 
goes on.
Dissipation of balls at the boundary, as in the 
sandpile\cite{btw}, can also be considered to allow the system 
to relax to a stationary state. Actually, in order to keep
the same definition of the model both in finite dimensions 
and in the random neighbor version, we consider here ``bulk
dissipation'' modifying step {\em 2)} into: 
{\em 2')} with probability $\lambda$ all white balls disappear,
otherwise {\em 2)} applies.

Let $e_i$ be the number of subsequent additions 
of white balls in urn $i$, since the last draw of a black one. 
Then urn $i$ contains $b$ black balls and 
$e_i+1$ white ones. The 
toppling probability, i.e.\ the probability to 
draw a black ball from urn $i$, is then
\beq
f_i=\frac{b}{e_i+b+1}.
\label{we1}
\eeq
In general, the probability that urn $i$ topples,
if it receives $q$ white balls from neighbors, is 
\beq
w_i(q)=
1-\frac{(e_i+q)!\Gamma(e_i+b+1)}{e_i!\Gamma(e_i+q+b+1)}\, .
\label{weq}
\eeq
The $\Gamma$-function is introduced here in order to generalize our 
discussion to any real positive $b$.

\begin{figure}
\resizebox{0.6\textwidth}{!}{%
  \includegraphics{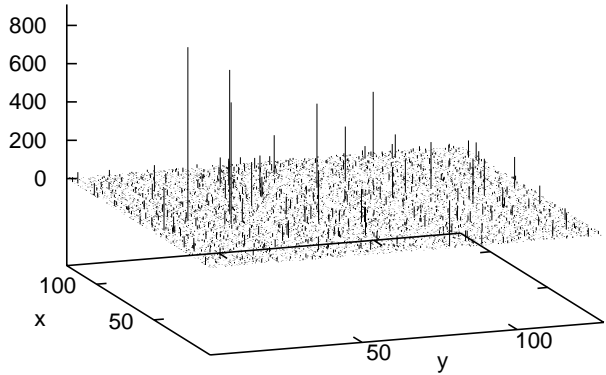}
}
\caption{Snapshot of a section of the $d=2$ system of size
 $L=128$ for $b=2.50$ and $\lambda=10^{-2}$.}
\label{fig1}
\end{figure}

The system spontaneously evolves to a critical
state which is generally characterized by a non uniform 
distribution of the variables $e_i$. A snapshot of the system 
for $b=2.5$ is shown in Fig.\ \ref{fig1}. It is clearly seen 
that very stable urns $e_i\gg 1$ coexist with less
stable ones. 

We say that the initial perturbation causes
an {\em avalanche} of size $s$ and volume $v$, 
where $s$ is the number of toppling events occurring 
before the system returns stable and $v$ is the number
of distinct sites involved in the avalanche. 
Operationally all urns which become unstable
after the first event topple simultaneously.
The urns, which as a result of this first
wave of topplings become unstable, topple 
simultaneously in a second wave, and so on. 
An avalanche is then also characterized by the 
number $w$ of such waves occurring before the 
system returns stable. 
Finally, one can also measure the total number
$E$ of white balls involved in the avalanche event.
Of course, if the initial urn resists 
absorbing an extra white ball, $s=v=w=E=0$. 
The distribution of avalanche sizes in the critical 
state
\beq
P(s)\sim s^{-\tau}
\label{deftau}
\eeq 
has a power law behavior. 
In this state the volume, waves and the number $E$ of the 
avalanche are related to its size by power laws 
\beq
v\sim s^{d/z},~~w\sim s^\omega,~~E\sim s^\gamma.
\label{defexp}
\eeq
The four exponents $\tau,~z,~\omega$ and $\gamma$
define the critical state of the model. They are given
in Table \ref{tab1} for several values of $b>1$
in $d=1$, $d=2$ and in the random neighbors version 
($K$ ``neighbors'' are randomly chosen
each time among all sites) respectively\footnote{The 
parameter $\lambda$, which introduces 
bulk dissipation, sets the upper cutoff of the critical
states. We found that, in $d=1,2$, 
dissipation at the boundaries, as in the sandpile
model, leads to similar results.
In the random neighbors version, we have found that the 
exponents do not depend on the number $K$ 
of neighbors chosen.}.
For $b\le 1$ the system becomes non-stationary:
The number of white balls in the system increases
linearly in time. In spite of this, we have found that
avalanches have still a well defined distribution.
We have analyzed in particular the border-line case $b=1$:
For the random neighbor model, 
we have found that in a range of times (from $10^5$ to
$2\cdot 10^8$), for various system sizes (up to
$n=2^{14}$ urns) and different dissipation 
rates ($\lambda=10^{-2}$ and $\lambda=10^{-3}$),
the distribution of avalanche sizes follows Eq.\ (\ref{deftau})
on more than three decades, with an exponent which
is, for all these cases, always in the range 
 $\tau\in[1.95,2.05]$. In $d=2$, for sizes up to $n=2^{14}$
and in a range of times from $5\cdot 10^7$ to $3\cdot 10^8$,
we have similarly found $\tau\in[2.03,2.11]$. In $d=1$, 
extensive simulations over lattice lengths of $256$ and $512$
sites and on a range of times from $10^9$ to $6\cdot 10^9$,
we have found $\tau\in[2.10,2.21]$.

\begin{table*}
\begin{tabular}{c|cccc|cccc|cc}
 & \multicolumn{4}{c|}{$d=1$} 
 & \multicolumn{4}{c|}{$d=2$} & \multicolumn{2}{c}{$d=\infty$} \\
 \hline
{$b$}
 & \multicolumn{1}{c}{$\tau$} & \multicolumn{1}{c}{$z$}
 & \multicolumn{1}{c}{$\omega$} 
 & \multicolumn{1}{c|}{$\gamma$} 
 & \multicolumn{1}{c}{$\tau$} & \multicolumn{1}{c}{$z$}
 & \multicolumn{1}{c}{$\omega$} 
 & \multicolumn{1}{c|}{$\gamma$} 
 & \multicolumn{1}{c}{$\tau$} & \multicolumn{1}{c}{$\omega$} \\
 \hline
$1.40$ & $1.80(2)$ & $2.45(1)$ & $0.61(1)$ & $1.44(1)$ 
       & $1.83(1)$ & $3.17(2)$ & $0.44(1)$ & $1.248(3)$
       & $1.885(7)$ & $0.34(3)$ \\
$1.60$ & $1.69(2)$ & $2.35(1)$ & $0.60(1) $ & $1.36(1)$ 
       & $1.74(1)$ & $3.10(2)$ & $0.45(1)$ & $1.183(3)$ 
       & $1.80(1)$ & $0.38(2) $ \\
$1.80$ & $1.61(2)$ & $2.29(1)$ & $0.60(1)$ & $1.31(1)$ 
       & $1.68(1)$ & $3.03(2)$ & $0.45(1)$ & $1.134(1)$
       & $1.68(1)$ & $0.40(2)$ \\
$2.00$ & $1.55(2)$ & $2.24(1)$ & $0.60(1)$ & $1.27(1)$ 
       & $1.63(1)$ & $2.97(2)$ & $0.46(1)$ & $1.103(1)$
       & $1.667(7)$ & $0.42(1)$ \\
$2.30$ & $1.47(2)$ & $2.18(1)$ & $0.59(1)$ & $1.23(1)$ 
       & $1.57(1)$ & $2.90(2)$ & $0.48(1)$ & $1.069(1)$
       & $1.642(9)$ & $0.44(1)$ \\
$2.50$ & $1.43(2)$ & $2.15(1)$ & $0.59(1)$ & $1.21(1)$ 
       & $1.55(1)$ & $2.84(2)$ & $0.49(1)$ & $1.056(1)$
       & $1.624(8)$ & $0.45(1)$ \\ 
$2.70$ & $1.39(2)$ & $2.13(1)$ & $0.59(1)$ & $1.19(1)$ 
       & $1.54(1)$ & $2.81(2)$ & $0.50(1)$ & $1.045(2)$
       & $1.609(8)$ & $0.47(1)$ \\
$3.00$ & $1.36(2)$ & $2.11(1)$ & $0.59(1)$ & $1.17(1)$ 
       & $1.52(1)$ & $2.76(2)$ & $0.51(1)$ & $1.034(3)$
       & $1.595(7)$ & $0.47(1)$ \\
\end{tabular}
\caption{The exponents for different values of $b$ for 
$d=1$ and sizes up to $L=512$, $d=2$ and $L=128$, and for 
the random neighbor model with $n=16384$ sites and $K=2$.}
\label{tab1}
\end{table*}

In an ecosystem species are 
probed by changes in the environment in a fashion which goes under the name of
coevolution. Namely, the environment, which is constituted by the interaction
among all living organisms, is implicitly modified by each single being 
and determines its fate as well\footnote{Just as an example, 
one could think at the 
selection pressure produced by the precence of oxygen in the air, 
which has been initially produced by some
zoo-phyte.}.
{}From the point of view of any single species, 
such perturbations impose a random selective pressure which eventually leads
to a change in the constitution of the species or to its extinction. 
A particular character, developed 
by a random mutation in a subspecies, can be ``selected''
by evolution if it becomes essential for the survival of 
the whole species in a changed environment \cite{Gould89}. 
In this process, which is driven by chance, species
become more and more complex. A more complex species
has also a wider variability in subspecies which,
in this process, will make it more resistant.
In much the same way, Polya urns in the model ``evolve''
increasing their ``complexity'' and their robustness to 
external perturbations. The model also 
assumes that the extinction of well adapted ``species'' 
(i.e. urns with $e_i\gg 1$) produces a larger perturbation
than that caused by poorly fit species.

In this toy ecosystem, each species has exactly the same 
chances of any other species to survive, 
when it appears ($e_i=0$). There is no {\em a priori}
genetic characteristic, such as fitness, which guarantees 
the survival of a species. Its survival will rather 
depend on its ``ability'' to adapt constantly, {\em via}
random mutations, to the changing environment.
This perspective, also suggests that an operational
measure of fitness (or resistance) of a species,
is possible using Eq.\ \req{we1}: High $e_i$ means
high fitness. Note that this differs from
the concept of fitness introduced in most SOC models 
of coevolution\cite{bs,modbs,netw}. In these, fitness
is related to reproduction rates, whereas in our 
simplified picture of coevolution,
fitness emerges as a measure of the resistance 
against extinction of a species. 

We show now that this different notion of fitness,
when introduced as an {\em intrinsic} property of
each species, leads to exaclty the same coevolutionary
process. To be more precise, let us assume that the 
probability $f_i$ of extinction of species $i$ 
under a perturbation, is no more given by Eq.\ \req{we1},
but it is rather fixed for each species.
In particular, this {\em intrinsic} property
is drawn randomly for each species from 
a distribution $\rho(f)$. Accordingly we also 
replace Eq.\ \req{weq} by $w_i(q)=1-(1-f_i)^q$. 
Species are probed by random perturbations (addition
of white balls) and, as before, the extinction of species 
$i$ perturbs ``neighboring'' species in the interaction 
web via the same toppling mechanism.
When a species disappears, its ``niche'' (site) is immediately 
occupied by a new species, with a new randomly drawn 
fitness value $f_i'$. Thus, for
\beq
\rho(f)=b f^{b-1}
\label{rhof}
\eeq 
we obtain exactly the same stochastic dynamics given by Eqs.\
(\ref{we1},\ref{weq}).  
In order to show the equivalence, it is enough
to show that Eq.\ \req{rhof} leads to the same rates
$w_i(q)$ of Eq.\ \req{weq}. Consider one particular urn and let
$e_{i,t}$ be the value of the corresponding variable after $t$
drawings. The event $e_{i,t}=e$ occurs with a probability
\footnote{Here 
$t$ is a ``local'' time on site
$i$, which measures the number of perturbations on that
site. Therefore the event $e_{i,t}=e$
implies $e_{i,t-e}=0$. Accordingly we used the
notation $P(e_{i,t}=e|e_{i,t-e}=0,f_i)$ (note indeed
that $P(e_{i,t}=0|e_{i,t}=0,f_i)=1$ for $e=0$).}
$P(e_{i,t}=e|e_{i,t-e}=0,f_i)=(1-f_i)^e$.
Taking the average over $\rho(f)$, we find
\beq
P(e_{i,t}=e|e_{i,t-e}=0)
=\frac{\Gamma(b+1)e!}
{\Gamma(e+b+1)}.
\label{Pe}
\eeq
Using detailed balance, $P(e_{i,t+q}=e+q|e_{i,t-e}=0)
=w_i(q)P(e_{i,t}=e|e_{i,t-e}=0)$, we easily recover
Eqs.\ (\ref{we1}, \ref{weq}).
The equivalence of the two models has been also tested 
in numerical simulations.

The initial definition of the model is completely
stochastic, whereas in the formulation based on Eq.\ 
\req{rhof}, $f_i$ are fixed, quenched 
variables, which are renewed stochastically
at each extinction event.
The equivalence of the two descriptions is an example 
of a general mapping\cite{RTS}, recently developed 
to deal with extremal dynamics. Its application in the
biological context are also discussed in 
Ref.\ \cite{memo}. 

There is a further interesting mapping, originally
developed in the context of interface growth\cite{mapping}, 
which can be applied to the present model. 
As in the sandpile model\cite{btw}, we define the toppling probability as 
\begin{eqnarray}
f_i=0\,\, \mbox{ if }\,\, e_i<h_i &
 \mbox{ and } & f_i=1\,\, \mbox{ if } \,\,e_i\ge h_i\,\, .
\label{thresh}
\end{eqnarray}
While in the sandpile the thresholds are fixed $h_i=2d$,
we introduce here a model where $h_i$ are randomly 
drawn from a given distribution $\psi(h)$ after each
toppling event on site $i$. The choice 
\beq
\psi(h)=\frac{b\Gamma(b+1)\Gamma(h)}{\Gamma(h+b+1)}
\label{psih}
\eeq
reproduces a dynamics which is equivalent to the
previous two formulations of the model.
To prove this, it is enough to derive the statistics
of the number $e_i=h_i-1$ of perturbations that an urn
with $f_i=f$ will overcome before toppling.
Clearly $P(h_i=h|f_i=f)=f (1-f)^{h-1}$. Taking the average
over the distribution \req{rhof} of $f$ leads indeed to
Eq.\ \req{psih}.

This formulation is completely deterministic: It
assumes that each urn appears with a prescribed ``lifetime'' 
measured in terms of perturbations. As soon as this 
lifetime is reached, the urn topples. This is the
same threshold dynamics used in the sandpile model.
Here however thresholds $h_i$ are very broadly
distributed (note that $\psi(h)\sim h^{-b-1}$),
whereas in the sandpile $\psi(h)=\delta(h-2d)$.
The sandpile is a paradigm for seismic phenomena:
Each site represent a fault which is perturbed 
by the slow addition of stress energy. When the
energy load $e_i$ of a site exceeds the threshold 
$h_i$, the fault breaks down and all the energy 
is released to neighbor sites. Our model also
proposes a different description of the same
phenomenon: each addition of stress energy
has the same probability $f_i$ to provoke 
an earthquake. When this occurs, it provokes 
the energy release and a local seismic rearrangement, 
i.e. $f_i\to f_i'$.

Using the quenched version of the model given by
Eq.\ (\ref{rhof}), it is easy to 
derive the effective distribution $\tilde\rho(f)$ of 
the $f_i$'s in the system at the stationary state.
It is enough to consider detailed balance in an
interval $f_i\in[f,f+df)$ under a single perturbation.
The probability that, in one time step, one $f_i$ leaves 
this interval is $f\tilde\rho(f) df$. This 
has to balance the number $\rho(f)df\int_0^1df'f'\tilde\rho(f')$
of sites which enter this interval.
This gives 
$\tilde\rho(f)=(b-1)f^{b-2}$. With Eq.\ \req{Pe}
one can also derive the distribution of $e_i$
in the system. Indeed, 
$P(e_{i,t}=e)=P(e_{i,t}=e|e_{i,t-e}=0) 
P(e_{i,t-e}=0)$ where $P(e_i=0)=1-1/b$ is derived 
imposing normalization. This leads to
$P(e_i)\sim e_i^{-b}$.

As $b\to 1^+$, both the distributions $\tilde\rho(f)$ 
and $P(e_i)$ become unnormalizable and the probability of
finding sites with $e_i=0$ vanishes. Accordingly, 
numerical simulations show that, for $b\le 1$, the
system average of $e_i$ increases linearly with time
and the system never reaches a stationary state.
The divergence of normalization of $\tilde\rho(f)$ 
at $f=0$ occurs because less ``fit'' species are more 
rapidly replaced than more ``fit'' ones. For $b\le 1$ 
the search for the ``perfect'' species $f_i=0$ never stops.
For $b>1$, the probability that a perturbation causes 
a toppling is 
$P_{\rm top}=\int_0^1 df f\tilde\rho(f)=\frac{b-1}{b}$,
which also vanishes as $b\to 1$. This means that, for
$b\le 1$, avalanches occur more and more 
rarely as time goes on.

Let us discuss in more detail the random neighbor
model. Numerical results are consistent with 
$z/D=1$ and $\gamma=\max[1,1/(b-1)]$, which is what 
one expects from the observation that each site is 
involved at most once in the same avalanche
(the exponent $\gamma=1/(b-1)$ for $1<b<2$ comes 
from the limit laws of Levy variables). 
On the other hand we see that the
exponents $\tau$ and $\omega$ differ from their usual
mean-field values $\tau=3/2$ and $\omega=1/2$, and that 
such values are eventually reached for $b\to\infty$.
This deviation $\tau\not = 3/2$ can be
understood as an effect of correlation.
In order to show this, let us review the argument
leading to $\tau=3/2$. Consider an avalanche and let 
$M_t$ be the number of unstable sites after $t$ 
toppling events. $M_t$ performs a random walk and
the size $s=\min\{t:~M_t=0,~t>0\}$ of the avalanche 
is the first return time of $M_t$ to $0$. 
Therefore $s$ has the same distribution
of the first return times to the origin of a random walk
$P(s)\sim s^{-3/2}$.
Since the steps $|M_t-M_{t-1}|$ of the random
walk are bounded by the coordination number $K$, 
the only possibility for a deviation from $\tau=3/2$ is 
to have correlations. 
Correlations indeed arise because a toppling event may 
release many white balls and these are 
transported along the avalanche thus modifying 
the probability of toppling of successive sites. 
The theoretical calculation of the exponents 
for the random neighbor version
is a challenging problem under current investigation.

In conclusion, we have introduced a very simple model 
which displays non-trivial self organized critical features.
The model was analyzed numerically and we also derived some 
analytic result. The main features of the model are 
non-homogeneous critical states, a critical non-stationary
state in a region of the control parameter ($b\le 1$) and 
non-trivial exponents even in the mean field limit. 
Furthermore, it allows for three different equivalent formulations,
which allow one to better investigate and understand the
nature of the critical state. As a closing remark, we notice
that the non-stationary regime can be relevant both for the 
description of ecological systems and for earthquake
dynamics. Both these systems are indeed not
in a stationary state. Interestingly enough, the exponent 
 $\tau$ is larger than $2$ for $b<1$. Both 
the Gutenberg-Richter law for earthquakes\cite{earthq} and 
the extinction size distribution\cite{fossil} also display 
avalanche exponents close to $2$.

\end{document}